\documentclass[fleqn,10pt]{wlscirep}
\usepackage[utf8]{inputenc}
\usepackage[T1]{fontenc}
\usepackage{booktabs}
\newcommand{\eg}{e.\,g.\ }
\newcommand{\ie}{i.\,e.\ }

\newcommand{\px}[2]{#1$\times$#2\,px}

\newenvironment{texttable}[2][1]
  {\def\arraystretch{#1}\tabular{#2}}
  {\endtabular}
 
\newcommand\Tstrut{\rule{0pt}{2.8ex}}

\newcommand{\appendixtitle}[2]{
    \begin{center}
    \textsc{Appendix #1}\\
    \textsc{#2}
    \end{center}
    }

\usepackage{placeins}
\usepackage{hyperref}

\title{Resolving challenges in deep learning-based analyses of histopathological images using explanation methods}

\author[1]{Miriam H\"agele}
\author[1]{Philipp Seegerer}
\author[2]{Sebastian Lapuschkin}
\author[3,4]{Michael Bockmayr}
\author[2]{Wojciech Samek}
\author[3*]{Frederick Klauschen}
\author[1,5,6*]{Klaus-Robert M\"uller}
\author[7*]{Alexander Binder}
\affil[1]{TU Berlin, Machine Learning Group, Berlin, 10587, Germany}
\affil[2]{Fraunhofer Heinrich Hertz Institute, Department of Video Coding \& Analytics, Berlin, 10587, Germany}
\affil[3]{Charit\'e University Hospital, Institute of Pathology, Berlin, 10117, Germany}
\affil[4]{University Medical Center Hamburg-Eppendorf, Department of Pediatric Hematology and Oncology, Hamburg, 20251, Germany}
\affil[5]{Korea University, Department of Brain and Cognitive Engineering, Anam-dong, Seongbuk-gu, Seoul, 02841, Korea}
\affil[6]{Max-Planck-Institute for Informatics, Campus E1 4, Saarbr\"ucken, 66123, Germany}
\affil[7]{Singapore University of Technology and Design, ISTD Pillar, Singapore, 487372, Singapore}

\affil[*]{frederick.klauschen@charite.de, klaus-robert.mueller@tu-berlin.de, alexander\_binder@sutd.edu.sg}


\begin{abstract}
Deep learning has recently gained popularity in digital pathology due to its high prediction quality. However, the medical domain requires explanation and insight for a better understanding beyond standard quantitative performance evaluation. Recently, many explanation methods have emerged. This work shows how heatmaps generated by these explanation methods allow to resolve common challenges encountered in deep learning-based digital histopathology analyses. We elaborate on biases which are typically inherent in histopathological image data. 
In the binary classification task of tumour tissue discrimination in publicly available haematoxylin-eosin-stained images of various tumour entities, we investigate three types of biases: (1) biases which affect the entire dataset, (2) biases which are by chance correlated with class labels and (3) sampling biases.
While standard analyses focus on patch-level evaluation, we advocate pixel-wise heatmaps, which offer a more precise and versatile diagnostic instrument. This insight is shown to not only be helpful to detect but also to remove the effects of common hidden biases, which improves generalisation within and across datasets. For example, we could see a trend of improved area under the receiver operating characteristic (ROC) curve by 5\% when reducing a labelling bias. 
Explanation techniques are thus demonstrated to be a helpful and highly relevant tool for the development and the deployment phases within the life cycle of real-world applications in digital pathology.
\end{abstract}
\begin{document}

\flushbottom
\maketitle

\thispagestyle{empty}

\section*{Introduction}
\label{sec:introduction}
With radiology and pathology, two central medical disciplines rely on imaging data for diagnostics. Histopathological diagnoses are made by trained experts by visually evaluating properties of the tissue on the microscopic level. However, the increasing diagnostic throughput and the need for additional quantitative assessments of tissue properties call for supportive computational image analysis approaches. Here, deep learning techniques offer solutions due to recent developments in increased predictive performance. But while predictive performance quality is essential, rendering the classification process transparent is crucial particularly for the medical application.\\
Recently, deep learning methods~\cite{lecun2015deep, schmidhuber2015deep} have successfully contributed among others to computer vision~\cite{szegedy2015going, krizhevsky2012imagenet}, where a broad variety of highly complex tasks have been studied. Learning methods in general and deep convolutional networks in particular have also received increased attention in medical imaging and digital pathology~\cite{litjens2016deep, litjens2017survey, ronneberger2015unet}. Notably, for the specific task of classifying skin lesions convolutional neural networks even show performance on par with medical professionals~\cite{esteva2017dermatologist}. These encouraging results are possible because of the non-linear structure of learning methods, which allows to model very complex high-dimensional functions. 
However, so far this high complexity comes at a cost: The clinical applicability of these learning machines is hampered by the fact that neither data scientists nor medical professionals comprehend their decision processes. High prediction capabilities without explanation of the prediction process itself is therefore considered suboptimal~\cite{lapuschkin2016analyzing, lapuschkin2019hans}. 
Only recently, explanation methods have been developed to analyse model decisions and introduce transparency for non-linear machine learning methods. Several visual explanation methods have been successfully applied to natural images in computer vision~\cite{simonyan2014deep, zeiler2014visualizing, yosinksi2015understanding, bach2015pixel, selvaraju2016grad, journals/corr/KindermansSAMD17, montavon2017explaining, zintgraf2017visualizing}.
This also holds high promise for medical imaging. Following the spirit of earlier work~\cite{binder2018towards,korbar2017looking}, we will study the usage of explanation methods specifically for addressing the challenges in digital pathology~\cite{Fuchs2011, Holzinger2017}. These challenges include (1) noisy ground truth labels due to intra- and inter-observer variability of pathologists, (2) stain variance across datasets, (3) highly imbalanced classes and (4) often only limited availability of labelled data. More importantly, these challenges make datasets prone to inherit various types of latent biases, which need to be addressed carefully in order to avoid encodings of these biases and other artefactual structures in the model. This is in particular true for digital pathology which is a highly interdisciplinary field, where medical and machine learning professionals need to collaborate closely. While labels and diagnoses are provided by highly skilled medical experts, typically involved machine learning experts only have basic knowledge of the medical domain. This gives rise to a non-negligible risk of unknowingly introducing biases. In clinical diagnostics, false predictions come at a high cost. Therefore it is absolutely essential to detect and remove hidden biases in the data that do not generalise to the generic patient case. \\
In this work, we contribute by introducing explainable neural networks into a digital pathology workflow. Explanation methods (such as Layer-wise Relevance Propagation (LRP)~\cite{bach2015pixel}) are able to generate visual explanations in the form of high-resolution heatmaps for single input images (cf. Fig.~\ref{fig:diff_entities}). Exemplarily, we demonstrate this for the detection of tumour tissue in haematoxylin-eosin-stained (H\&E) slides of different tumour entities, namely skin melanoma (TCGA-Skin Cutaneous Melanoma), breast carcinoma (TCGA-Breast Invasive Carcinoma) and lung carcinoma (TCGA-Lung Adenocarcinoma) using LRP. Such heatmaps allow to detect latent or unknown biases on single affected images, often without the necessity of labels. Therefore, they provide a significant advantage over standard performance metrics in a field where labelled data is a time-consuming, experts-requiring process. Furthermore, rendered heatmaps allow to identify whether the features employed by neural networks are consistent with medical insight of domain experts, both qualitatively and quantitatively. To quantify the performance in a well-controlled manner, we evaluate our model on the relevant scale of cells instead of patches.
The five performed experiments, which demonstrate the benefits of explainable methods in terms of cell level evaluation as well as uncovering hidden dataset biases, are listed in Tab.~\ref{tab:experiments}. Finally, the benefits of visual explanations, the advantages of pixel-wise over coarse heatmaps and the limitations of explanation methods in digital pathology are discussed and put into perspective.

\section*{Related work}
\label{sec:related_work}

\subsection*{Models in digital pathology}
Similar to the recent trend in computer vision, where end-to-end training with deep learning clearly prevails classification of handcrafted features, an increased use of deep learning, e.g. convolutional neural networks (CNN) is also noticeable in digital pathology. Nonetheless, there are some works on combining support vector machines with image feature algorithms \cite{journals/artmed/Cruz-RoaCG11, huang2017SVM, binder2018towards, yuan2012quantitative}. 
In deep learning several works propose custom-designed networks~\cite{sirinukunwattana2016locality,xu2016stacked}, \eg spatially constrained, locality sensitive CNNs for the detection of nuclei in histopathological images~\cite{sirinukunwattana2016locality}, while most often standard deep learning architectures (\eg AlexNet~\cite{krizhevsky2012imagenet}, GoogLeNet~\cite{szegedy2015going}, ResNet~\cite{he2016deep}) as well as hybrids are used for digital pathology~\cite{xu2017large, wang2016deep, korbar2017looking, khosravi2017deep, bejnordi2017context}. According to Litjens et al.~\cite{litjens2017survey}, the GoogLeNet Inception-V3 model is currently the most commonly used architecture in digital pathology.

\subsection*{Interpretability in computational pathology}
As discussed above, more and more developments have emerged that introduce the possibility of explanation~\cite{simonyan2014deep, zeiler2014visualizing, yosinksi2015understanding, bach2015pixel, selvaraju2016grad, journals/corr/KindermansSAMD17, montavon2017explaining, zintgraf2017visualizing} (for a summary of implementations cf. Alber et al.~\cite{alber2019innvestigate}); few of which have been applied in digital pathology~\cite{cruz2013deep, xu2017large, korbar2017looking, binder2018towards, klauschen2018scoring, graziani2018regression, mobadersany2018predicting, liu17, coudray2018classification}. 
The visualisation of a support vector machine's decision on Bag-of-Visual-Words features in a histopathological discrimination task is explored in Binder et al.~\cite{binder2018towards}. The authors present an explanatory approach for evidence of tumour and lymphocytes in H\&E images as well as for molecular properties which--unlike nuclei--are not visually apparent in histomorphological H\&E stains.
There have also been a few prior works on visualisation of deep learning-based models in digital pathology. For example Cruz-Roa et al.\ (2013) have extended a neural network by a so-called ``digital staining'' procedure~\cite{cruz2013deep}. This was achieved by weighting each convolutional feature map in the last hidden layer by the associated weight of the softmax classifier and combining those weighted feature maps into a single heatmap image.
Xu et al.\ (2017) visualised the response of neurons in the last hidden layer of a CNN for brain tumour classification and segmentation~\cite{xu2017large}.
Quite frequently, prediction probabilities are visualised to explain the model's decision on patch-level \cite{liu17, mobadersany2018predicting, coudray2018classification}. Liu et al.\ (2017) additionally report a localisation score based on free-response receiver operating characteristic analysis for their heatmaps~\cite{liu17}. Korbar et al.\ (2017) compare different gradient-based visualisation methods to highlight relevant structures in whole-slide H\&E images of colorectal polyps based on the classification of a deep residual neural network~\cite{korbar2017looking}. In order to compare different approaches, regions of interest are derived from the visualisation heatmaps. The quality of these regions is subsequently quantified by the agreement with expert-annotated regions.

\section*{Method}
\label{sec:method}

\subsection*{Datasets}
We demonstrate the versatility of our approach, in terms of tumour entities, by its application to three different projects of the TCGA Research Network\cite{TCGA}, namely cutaneous malignant melanoma (TCGA-Skin Cutaneous Melanoma, SKCM), invasive breast cancer (TCGA-Breast Invasive Carcinoma, BRCA) and lung adenocarcinoma (TCGA-Lung Adenocarcinoma, LUAD). All three projects consist of H\&E stained images originating from various study sites and represent a wide range of commonly encountered variability. For every project regions of interest were chosen from the corresponding whole-slide images and annotated by a board-certified pathologist. Annotation labels comprise normal, pathological and artefactual tissue components. Subsequently, image patches of size \px{200}{200} and corresponding binary labels (tumor vs. no tumor) were extracted from these large-area annotations (\url{github.com/MiriamHaegele/patch_tcga}). A summary of the available data for the different experiments can be found in Table~\ref{tab:data}. Note that both the total number of available patches and the patch ratio between classes vary highly between the datasets. Thus the data reflects the common medical dataset challenges of very limited amount of annotated data and high class imbalances. These imbalances are due to the very different distributions of specific cell types.\\
For quantitative evaluation of the models, individual cells of a representative held-out subset were annotated for each project. For this all cells from five tiles sized \px{1000}{1000} to \px{2000}{2000} (depending on tumour entity) were annotated by a board-certified pathologist from our team. Tiles were chosen to represent a wide range of histopathological tissues. In addition to cell annotations, we introduced area annotations for vessels, necrosis and artefacts. 
Cells that could not be clearly identified due to their surface cut or their ambiguity were excluded from the annotation. For a visual and quantitative summary of the extensively labelled test datasets, the reader is referred to Supplemental Fig.~2-4  and Supplemental Tab.~1, respectively. 
In summary, we annotated a total number of 1,803 (BRCA), 3,961 (SKCM) and 2,722 (LUAD) cells for quantitative evaluation.

\begin{table}[h]
	\centering
	\begin{tabular}{crrrc}
	\toprule
	 Tumour & \multicolumn{1}{c}{Number} & \multicolumn{1}{c}{Total number} & \multicolumn{1}{c}{Number of} & $F_1$-score\\
	 entity & \multicolumn{1}{c}{of cases} & \multicolumn{1}{c}{of patches} & \multicolumn{1}{c}{tumour patches} & (weighted) \\
	 \midrule
	 \Tstrut 
	 SKCM 			& 38 & 26,746   &  19,139\, (71.6 \%) & \textbf{0.87}\\
	 BRCA 			& 72 & 2,748    &  2,308\, (84.0 \%) & \textbf{0.95}\\
	 LUAD 			& 39 & 13,165   &  4,805\, (36.5 \%) & \textbf{0.90}\\
	 \bottomrule
	\end{tabular}
	\caption{Summary of the available data. The classification performance per tumour entity is reported as weighted $F_1$-score. The $F1$-score is defined as the harmonic mean of precision and recall and ranges between 0 and 1. To account for class imbalances, we compute the weighted average of $F_1$-scores of both labels, weighted by their support.}
	\label{tab:data}
\end{table}

\subsection*{Convolutional neural network training}
We demonstrate the following analyses using the GoogLeNet architecture~\cite{szegedy2015going}, which is well established for generic computer vision tasks and which has recently been proven to also work well in digital pathology~\cite{litjens2017survey}. 
In particular, we finetuned a pretrained GoogLeNet from the \textit{Caffe Model Zoo}~\cite{jia2014caffe} (BAIR/BVLC GoogLeNet Model) for each tumour entity. We modified the GoogLeNet to process images of a non-standard input size by replacing the last average pooling layer by a global pooling layer which allows for an adjustable size of the pooling operation depending on the size of its input.
For the training and test set patient cases were split 80/20, while keeping the ratio of healthy and diseased cases constant. Patches were sampled from the corresponding cases and training patches were augmented with random translation and rotation.
Optimisation was performed via SGD with a learning rate of $10^{-3}$ and a learning rate schedule which reduced the learning rate by a factor of 10 every 10 epochs. Mini-batches comprised 128 images each. All other hyperparameters were set to their default values. Note that auxiliary classifiers of the GoogLeNet were not learned due to a lack of performance increase. In order to counteract class imbalances, we randomly oversampled the minority class in every mini-batch to reach uniform class distribution. Furthermore, we performed a \mbox{3-fold} cross validation to determine the epoch for early stopping (maximal number of epochs was set to 50), which was chosen as the lowest validation error averaged over all folds. This enabled us to make best use of the limited available data by using the full training set to train the model.
The models' performance is reported as $F_{1}$ scores. Additionally, we report the confusion matrices in the Supplement in order to avoid the dependency of single-valued performance measurements on class distributions.

\subsection*{Explaining classifier decisions}
Subsequent to training neural networks for the binary classification tasks, we used Layer-wise Relevance Propagation (LRP) to provide pixel-level explanation heatmaps for the classification decision~\cite{bach2015pixel}. LRP attributes the network's output to input dimensions by distributing the output value through a backward pass. Thereby a relevance score is assigned to every pixel, indicating how much the individual pixel contributes to the classifier decision in favour of a particular class. These high-resolution heatmaps are centred around zero and after normalisation lie within the range of [-1,1], where negative values (depicted in blue) contradict the particular class while positive values (depicted in red) are in favour of that class.\\
More specifically, in this work we applied a combination of the following LRP rules to distribute the relevance scores $R_{i}^{(l)}$ at neuron $i$ in layer $l$ to input space (similar to Lapuschkin et al.~\cite{lapuschkin2017understanding}). $z_{ij}^{(+/-)}$ denotes the (pos./ neg.) pre-activation of neuron $i$ in layer $l$ with connection to neuron $j$ in layer $l+1$.
Relevance is distributed from the network's output through the fully connected classification layer according to the $\epsilon$-rule ($\epsilon=1$):
\begin{equation}
    R_{i}^{(l)} = \sum_{j} \frac{z_{ij}}{\sum_{i'}z_{i'j}+ \epsilon \cdot \text{sign}(\sum_{i'}z_{i'j})} R_{j}^{(l+1)} \enspace .
\end{equation}
For the succeeding convolutional layers we applied the $\alpha \beta$-rule ($\alpha=1$ and $\beta=0$):
\begin{equation}
    R_{i}^{(l)} = \sum_{j} (\alpha \cdot \frac{z_{ij}^{+}}{\sum_{i'}z_{i'j}^{+}} + \beta \cdot \frac{z_{ij}^{-}}{\sum_{i'}z_{i'j}^{-}}) R_{j}^{(l+1)} \enspace .
\end{equation}
The choice of parameters is based on results from Montavon et al.~\cite{montavon2017methods}. \\
 Due to the fixed input size of neural networks, we need to aggregate heatmaps of this size in order to create visual explanation for larger tiles. Note that normalising the heatmap at the tile level (instead of normalising individual patches) enables us to preserve peaks and overall relation between relevance of the patches. This is desirable because the sum of relevance scores per patch represents the evidence of the classifier for a particular class. 
In order to allow for comparison between tile heatmaps, we take the global maximum of all computed heatmaps for normalisation instead of normalising on single tile level. All analyses are based on these settings.
For visualisation purposes (\eg Fig.~\ref{fig:diff_entities}), we used locally normalised, one-tenth overlapping patches. Furthermore, images were superimposed onto the grey-scaled H\&E stain.
For all our experiments, we made use of the publicly available Caffe implementation of LRP \cite{lapuschkin2016lrp}.

\subsection*{Quantitative evaluation of explanation heatmaps}
Going beyond visual inspection of the heatmaps, we evaluate the classifiers more thoroughly on the representative tiles, where ground truth annotations per cell were available. Region annotations are treated as single structures in this analysis. On the supposition that an optimal model, for the binary classification task of identifying tumour, will take into account all cancer cells, we computed a receiver operating characteristic (ROC) curve on normalised heatmaps. Hence, the ROC curve depicts the detection sensitivity of cancer cells in relation to the false positive rate for various thresholds of positive relevance, \ie relevance in favour of predicting class \emph{cancer}. Note that we only considered rectified heatmaps, \ie positive relevance values. To compute the ROC curve on cell level, we calculated a single relevance value for all labelled cells. Individual cells were assigned the mean value of a circular area around their corresponding point annotation. The acceptance radius around point annotations was set to half the diameter of an average cancer cell under the given resolution ($\sim$ 50 px; measured on a representative patch).
We additionally report the area under the curve (AUC), which is a metric particularly suited for highly imbalanced classes.

\section*{Experimental results}
\label{sec:experiments}

We conduct the following five experiments to study various beneficial aspects of including explanation heatmaps in deep learning-based medical imaging analyses (for a brief summary of experiments and their key messages see Table~\ref{tab:experiments}). Overall, two main insights can be found: (1) Explanation heatmaps do not only enable domain experts to visually inspect learned feature, but also allow for {\em quantitative} evaluations on cell level. (2) Besides, high-resolution heatmaps help to reveal and aid in subsequently removal of various types of latent biases in the data.

\begin{table*}[h!]
	\footnotesize
	\centering
    \begin{texttable}{p{2.9cm}p{3cm}p{1cm}p{3.7cm}p{3.2cm}}
	\toprule
	\multicolumn{1}{c}{Experiments} & \multicolumn{1}{c}{Description} & \multicolumn{1}{c}{Heatmaps} & \multicolumn{1}{c}{Benefit of visual explanation} & \multicolumn{1}{c}{Lifecycle phase}\\
	\midrule
	\Tstrut 
    Feature visualisation & Tumour classification in                                                     various entities (BRCA, SKCM \& LUAD)
                                    &\mbox{Fig.~\ref{fig:diff_entities}} 
                                    & Visual and quantitative verification of  learned features on cell level
                                    & Deployment phase (\eg computer-aided diagnosis systems)\\
    \rule{0pt}{5ex}
    Class sampling \mbox{ratios}           
                                    & Different class sampling ratios in           mini-batches
                                    & Fig.~\ref{fig:sampling}
                                    & Deliberate manipulation for different application use cases (contrary effects of recall and precision)
                                    & Deployment phase \\
    \rule{0pt}{5ex}                                
    Dataset bias 			        & Label bias affecting entire dataset
                                    & \mbox{Fig.~\ref{fig:bias_summary}} \mbox{(top)}
                                    & Bias detectable on a single sample, no additional held-out data necessary
                                    & Development phase \\
    \rule{0pt}{5ex}                                
    \mbox{"Class } correlated" bias 		
                                    & Artificial corruption correlated with one    class label
                                    & \mbox{Fig.~\ref{fig:bias_summary}} \mbox{(middle)}
                                    & Bias detectable on a single sample, possible to detect very small artefacts
                                    & Development phase \\
    \rule{0pt}{5ex}                                
    Sample bias                     & Exclusion of a tissue type in the                                               training data (here: necrosis)
                                    & \mbox{Fig.~\ref{fig:bias_summary}} \mbox{(bottom)}
                                    & Bias detectable on few samples of the missing tissue type, small regions of the missing tissue type also precisely detectable
                                    & Development phase (\ie iterative process to create comprehensive dataset) \\
                                
    \bottomrule
    \end{texttable}
    \caption{Brief summary of the performed experiments including their key messages.}
    \label{tab:experiments}
\end{table*}

\subsection*{Verifying learned features}
\label{sec:experiments_verifying_features}

\begin{figure}[h!]
	\centering
	\includegraphics[width=0.65\textwidth]{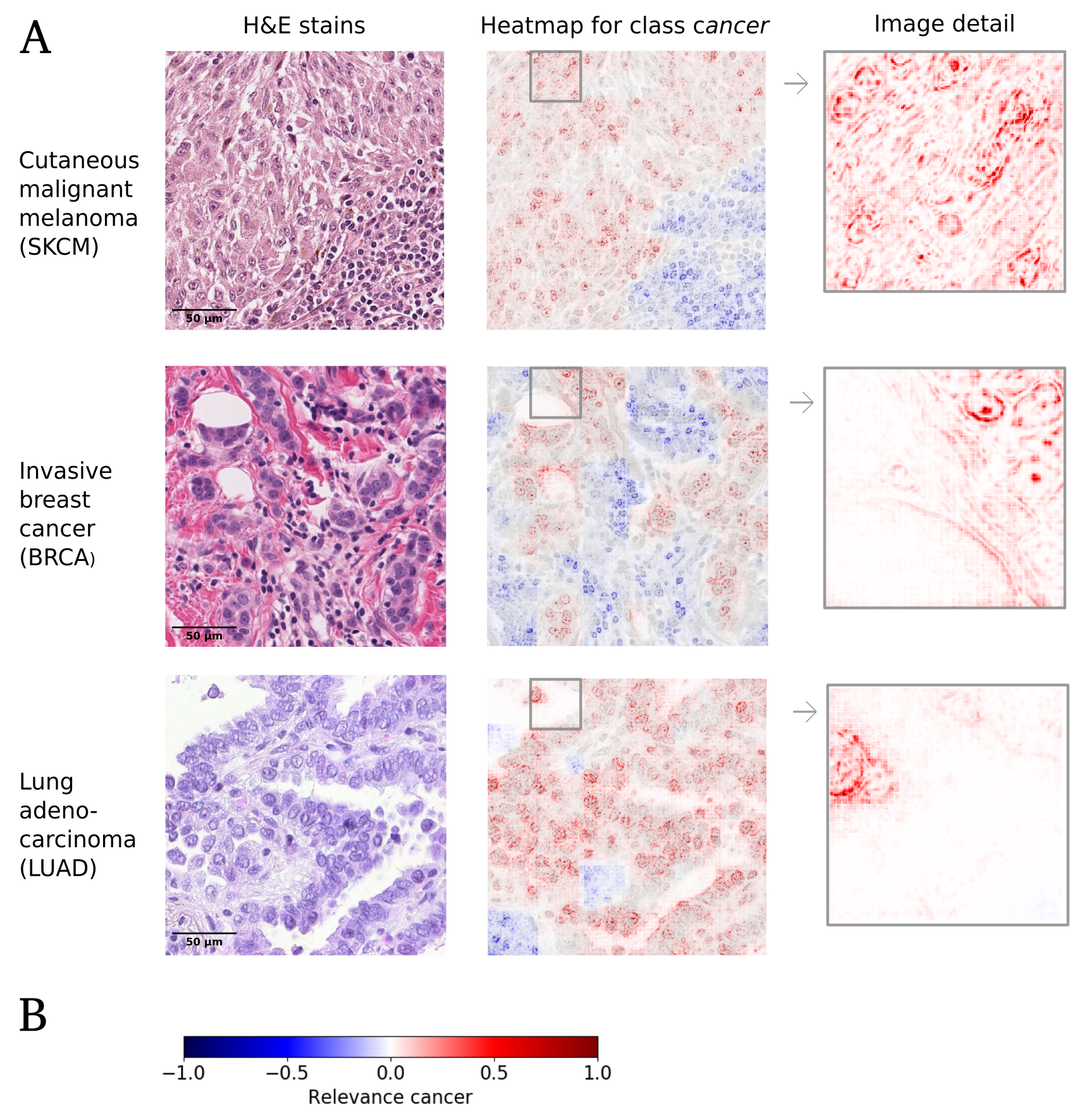}
	\caption{\textbf{A}. Exemplary visual explanations (superimposed on grey-scaled H\&E stain) for the three studied tumour entities. Red denotes positive relevance, \ie in favour for the prediction of class \emph{cancer}, while blue denotes negative relevance, \ie contradicting the prediction of class \emph{cancer}. Corresponding annotations can be found in Supplemental Fig.~3 (blue boxes). The image details illustrate the high-resolution of the computed heatmaps. \textbf{B}. Corresponding colourbar for the relevance distribution of cancerous tissue classification, which is used throughout the paper.}
	\label{fig:diff_entities}
\end{figure}

\begin{figure}[h]
	\centering
	\includegraphics[width=0.4\textwidth]{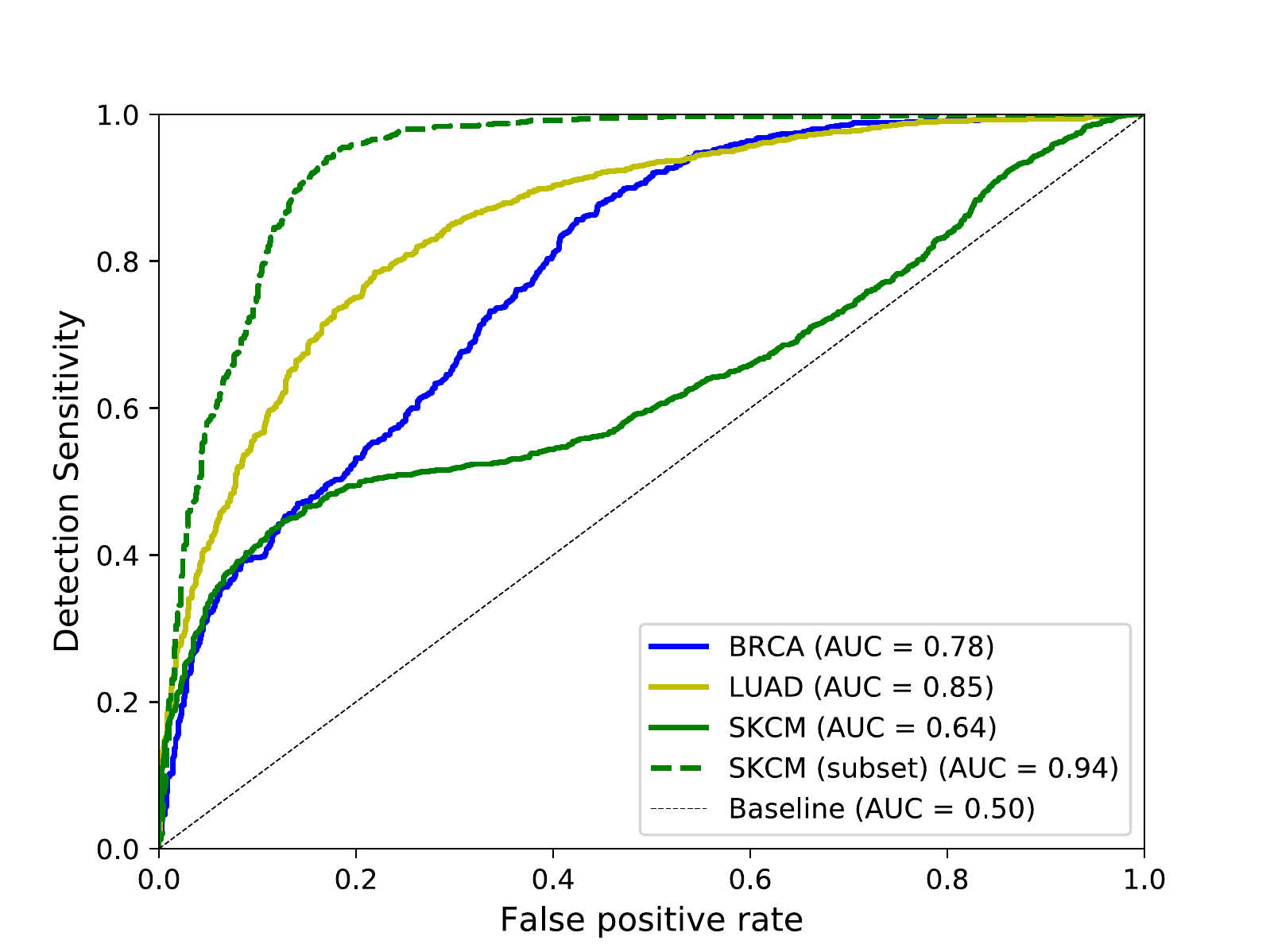}
    \caption{Quantitative evaluation of heatmaps: ROC curves for single cell recognition, quantified by their area under the curve (AUC). More specifically, we evaluate the presence of positive relevance on task specific structures (\ie cancer cells). For more details, please see Section \emph{Experiments (Verifying learned features)}.}
    \label{fig:ROC}
\end{figure}

For each of the three tumour entities we trained a separate neural network to discriminate between healthy and cancerous tissue. Performance is reported as weighted $F_1$-score which computes the harmonic mean between precision and recall for both classes and weights the average by the corresponding supports (i.e. number of true instances per label). The corresponding weighted $F_1$-scores for held-out patients range between 0.87 and 0.95 (cf. Tab.~\ref{tab:data}, for confusion matrices cf. Supplemental Fig.~1).
Subsequent to training the neural networks, explanation heatmaps of the respective extensively labelled tiles were computed. One exemplary heatmap per entity for class {\em cancer} is displayed in Fig.~\ref{fig:diff_entities}A. To make visual inspection easier, we overlayed heatmaps with their corresponding grey-scale version of the original stain.
The enlarged image detail illustrates the fine granularity of the visual explanations. Furthermore it becomes apparent from this image detail that high positive relevance values are spatially located on the nuclei and in particular on nucleoli and nuclear membrane as well as cytoplasm.
In order to report the performance on all chosen tiles, we computed ROC curves for single cell recognition and their corresponding AUCs. As baseline we choose heatmaps which solely consist of zeros or solely consist of ones (AUC=0.5). We additionally considered the case of randomly generated heatmaps, leading to AUCs of 0.5$\pm$ 0.0002 (over 100 runs). The results are displayed in Fig.~\ref{fig:ROC}.  For all three tumour entities the AUC clearly surpasses these baselines, \ie the computed heatmaps have a significant overlap with the labels provided by a domain expert.
We observe that the curves for BRCA and LUAD are of similar shape, exhibiting desired ROC properties as well as high AUC values. Considering the ROC curve of the SKCM classifier, it displays less accurate predictions per cell with regard to small relevance values. Taking a closer look at the SKCM heatmaps and the corresponding ROC curves of single tiles, we can trace this back to two tiles that contain tumour-infiltrating lymphocytes (cf. Supplemental Fig.~4). These also received small but positive relevance, leading to the observed flattening of the ROC curve in Fig.~\ref{fig:ROC}.
In summary, the observed heatmaps provide features that qualitatively and quantitatively show a high overlap to human labels and thus conform to histopathological knowledge.

\subsection*{Data sampling strategies}
\begin{figure}[h]
	\centering
    \includegraphics[width=0.5\textwidth]{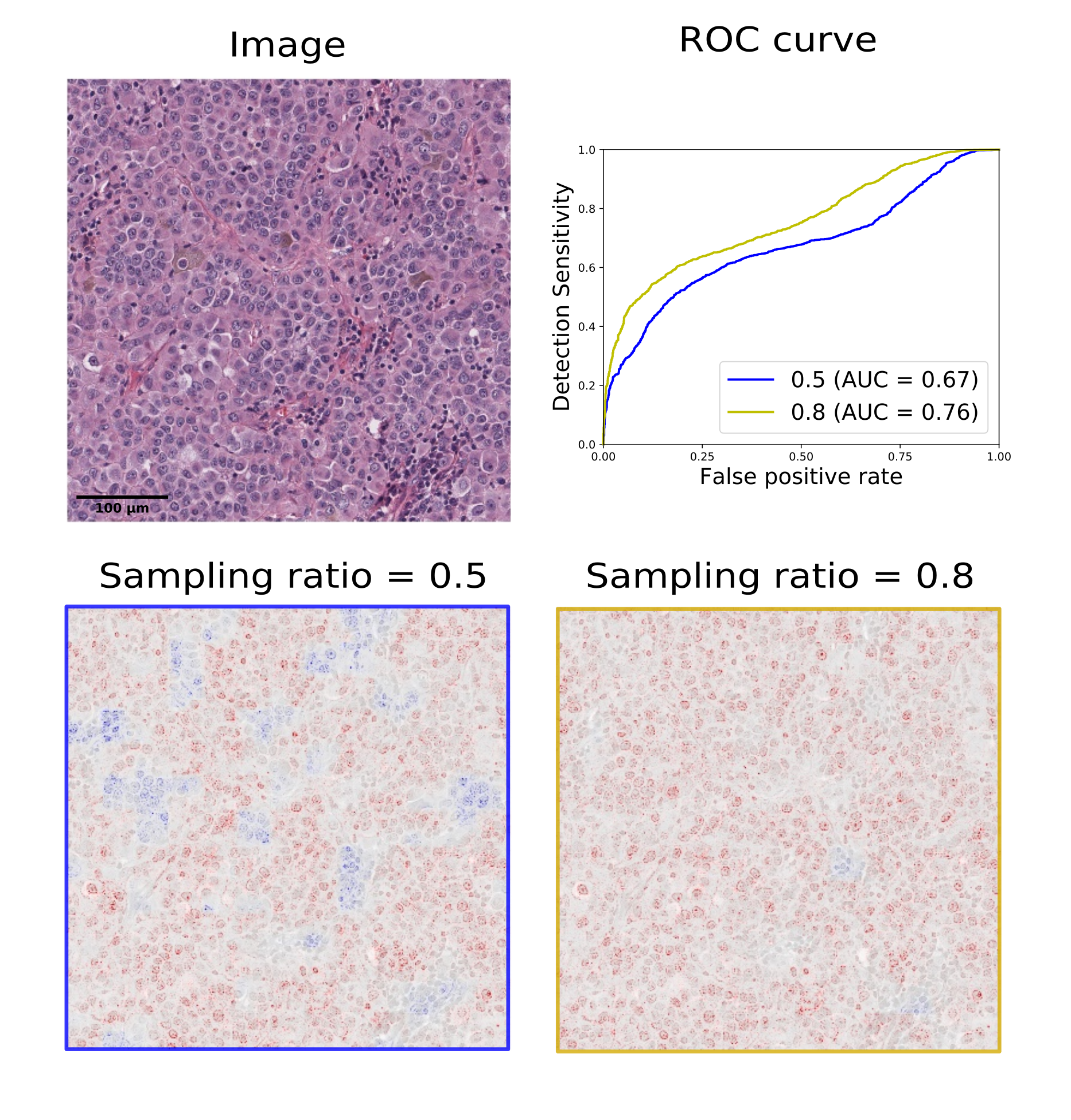}
	\caption{Investigating the influence of fixed class sampling ratios in mini-batches on explanation heatmaps, in particular, sampling ratios of 0.5 and 0.8 in favour of class \emph{cancer}. Heatmaps of the classifier with uniform sampling depicts positive and negative relevance (left) whereas the one of tumour tissue oversampling depicts almost only and slightly more positive relevance. Thus, the higher recall on patch level is also reflected in the heatmaps as additionally demonstrated quantitatively in the ROC curve.}
	\label{fig:sampling}
\end{figure}

The design of the training process is generically reflected in the model's ability to generalise and, as described in the following, it can also be observed in its heatmap. 
A possible solution to tackle the typical challenge of highly imbalanced classes in digital pathology is to define a fixed ratio of class labels in mini-batches during training, \eg by oversampling from the minority class. In this experiment, we explore the influence of different sampling ratios in terms of model performance and explanation heatmaps.
In particular, we inspect two scenarios, where firstly we simulate the balanced case by uniformly sampling from both classes and secondly we set the ratio to 0.8 in favour of the cancer class.\\
Despite their similar performance with regard to single-value accuracy measures such as $F_1$, we notice, as expected, a different trend in recall and precision. While the classifier that receives class-balanced samples in every mini-batch has higher recall of the positive class, the precision is slightly worse than in the uniform sampling case (cf. Supplemental Tab.~2). Heatmaps of these models reflect this observed trend by predominance of positive relevance in the cancer dominant scenario and a more balanced occurrence of positive and negative relevance in the uniform sampling case (cf. Fig.~\ref{fig:sampling}). The ROC curve over the exemplary heatmaps confirms this observed trend of increased recall in the cancerous tissue dominant scenario (cf. ROC curves in Fig.~\ref{fig:sampling}). While we demonstrate this effect for SKCM, similar trends could also be seen for BRCA and LUAD. 
In summary, heatmaps reflect the influence of fixed sampling ratios, which are known from common accuracy measures without the need to rely on fine-grainedly labelled testing data. They can thus be adapted to specific medical application purposes. 

\subsection*{Uncovering biases}
\label{sec:experiments_undesidered_strategies}
\begin{figure*}[h]
    \centering
    \includegraphics[width=0.9\textwidth]{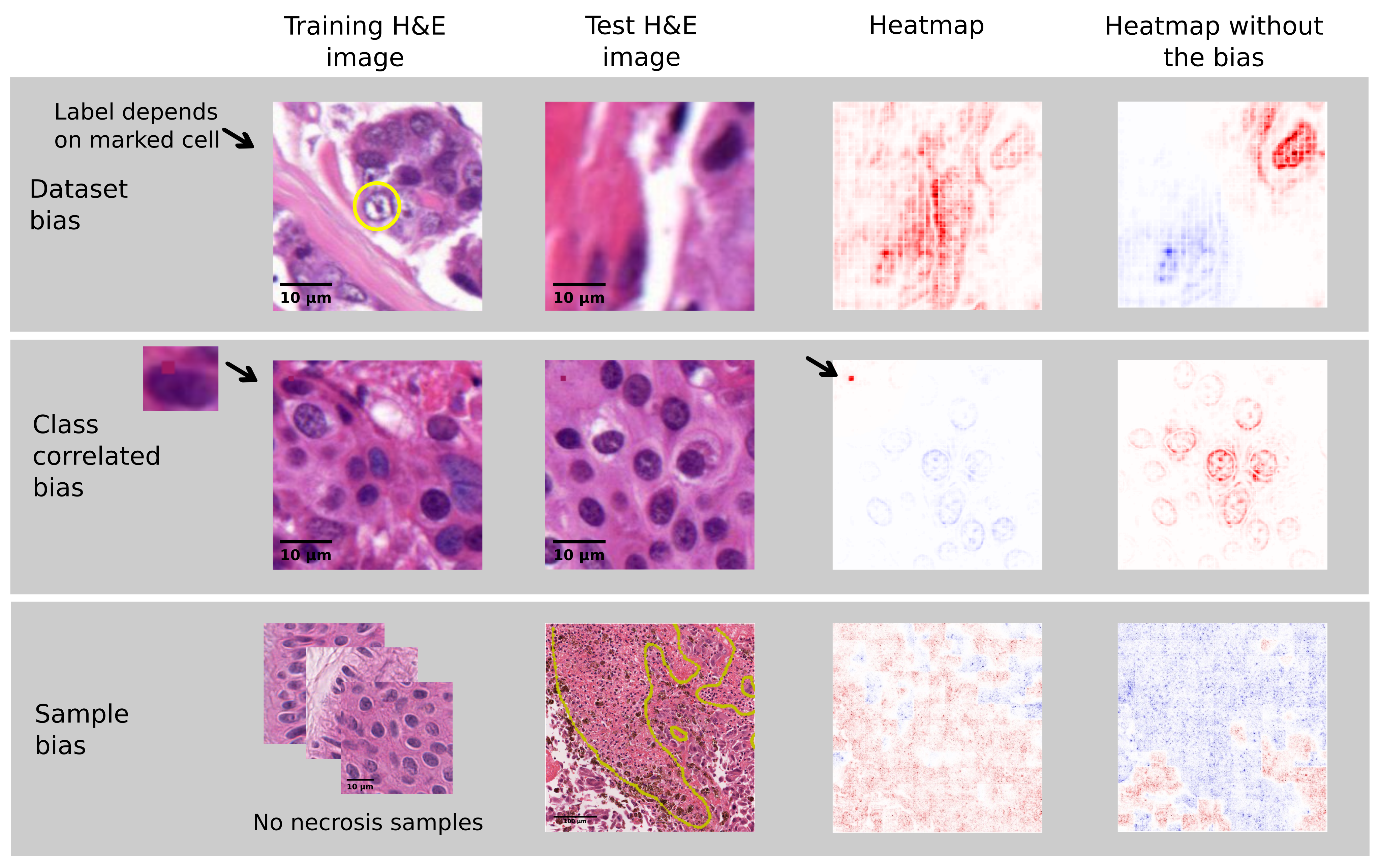}
	\caption{Illustration of the effects of the three studied types of biases on high-resolution explanation heatmaps. These are contrasted against heatmaps of models which are not affected by the biases (right). 
\textbf{Dataset bias}. This dataset is characterised by a label bias which results from determining the label solely from the patch's centre cell (yellow mark). The heatmap demonstrates how the network therefore learns to focus on the centre of the patch. 
\textbf{Class-correlated bias}. Detection of a class related bias in form of a small artificial corruption. The heatmap reveals that the model has based its decision on the bias instead of relevant biological features. High-resolution heatmaps are able to identify these class-correlated biases in a single example and to accurately pinpoint to even very small artefacts. 
\textbf{Sample bias}. Demonstrating the effect of sampling biases by training a classifier on a dataset lacking examples of necrosis.
The presented exemplary tile presents, apart from necrotic tissue, also cancer cells to show the correct classification of the target class in both cases while the assessment of necrotic tissue differs between the classifiers.}
	\label{fig:bias_summary}
\end{figure*}

In the following experiments, we elaborate on the potential of explanation heatmaps to uncover latent but crucial biases in the data. These biases can be crucial in the sense that they affect generalisation to unseen data and therefore limit the application of deep learning-based analyses of histopathological images to real-world scenarios. The greatest challenge with data biases is that they can often be indiscernible or at least hard to detect with common accuracy measures whereas high-resolution heatmaps allow to resolve these undesirable strategies on a single sample, affected by the bias. In the following, we study three different types of biases: (1) biases that affect the entire dataset, (2) biases that are by chance correlated with class labels and (3) sampling biases. An overview of the results can be found in Fig.~\ref{fig:bias_summary}. \\

\noindent \emph{Dataset bias.} \enspace
This first experiment is concerned with the detection of biases that affect the entire dataset, \ie biases that are spread across all classes, including test data. This can lead to an allegedly good test performance whereas performances on independent datasets, \ie cross-dataset generalisation, will be poor. This is due to the fact that the model might make use of the bias during training under the assumption that it is a characteristic trait of this image type. 
These biases often get introduced unintentionally in the process of creating datasets because of the unawareness that algorithms lack the human medical expert's ability to ignore clearly irrelevant noise. 
Here, we use a dataset where labels solely depend on the cell in the patch's centre independent of any other occurring cell types. The dataset comprises 2116 H\&E stained image tiles (\px{200}{200}) from a breast-cancer dataset \cite{binder2018towards} and provides single labels per patch.
As described above, we trained a neural network to discriminate between labels of cancer and healthy cells. In order to amplify the effect, we trained the network without including spatial translations in the data augmentation process. For purposes of comparison, we trained an additional network on random \px{120}{120} crops from the original dataset.\\
Contrasting the heatmaps of both models, as exemplarily depicted in the top row of Fig.~\ref{fig:bias_summary}, we can, as expected, observe a focus on the centre cell for the not translation invariant model. Therefore, the network sometimes fails to detect cancerous tissue in out-of-sample tiles---predominantly when cancer cells are not at the patch's centre---even though it allegedly performed well on the centred test data. Ensuring that this trend holds across the entire, held-out test dataset, we computed the mean over absolute relevance scores of all contained patches (cf. Supplemental Fig.~5). In order to quantify this effect, we consider rectangular areas around the patch's centre of different diameters (Supplemental Fig.~6). It becomes apparent that the absolute relevance, \ie independent of its sign, is more focused on the centre of the patch with less relevance towards the borders while it is more evenly spread out across the entire patch when the bias is counteracted with spatial translation augmentations. The AUC improves by 5\% (cf. Supplemental Fig.~7). \\

\noindent \emph{"Class-correlated" bias.} \enspace
The second type of biases, which we are considering in this work, are image features that are unintentionally correlated with labels in the dataset. The risk of this type of bias is significantly increased if the number of samples per class is rather small, since structural correlations can occur more easily. 
These biases can in principle occur from artefacts, which do not have any biological meaning or they can also occur from cell types or patterns that often but not necessarily always co-occur with the label.
We demonstrate the class-correlated bias under controlled conditions by artificially corrupting all patches of class cancer. More specifically, we replace a square region of size \px{5}{5} in the top left corner with a single colour that is close to the colour scheme of H\&E stained images. For some images this leads to hardly visible artefacts (cf. left image in middle row of Fig.~\ref{fig:bias_summary}). \\
Training a classifier on this corrupted dataset, led to a perfect test accuracy of 100\%. Visual inspection of the heatmaps, however, indicate that the classifier learned to focus on the artefact instead of biologically relevant features (cf. middle row Fig.~\ref{fig:bias_summary}). This finding persists when averaging over all heatmaps of a particular class (cf. Supplemental Fig.~10). This flaw can easily be detected on heatmaps of even a single affected patch from the (original) test data.\\

\noindent \emph{Sampling bias.} \enspace
In binary or multi-class classifications exhaustive datasets are of high importance in order to learn a representative manifold. Visual explanations of the decision process can help to detect dataset shifts in terms of missing structures of histopathological images, \ie certain cell types, biological or (staining) artefacts in the training data. Otherwise, at test time, patches might not lie on the learned manifold, leading to unpredictable behaviour of the classifier. This means that the training data should ideally be representative, containing all possible sample classes, independent of the classification task. Only then reliable discrimination between classes can be guaranteed. 
Neural networks are usually trained on only few region of interests of very complex whole slide images (WSI) with the aim to apply the model in a sliding window approach on WSIs at test time. Hence, it is necessary that the model is trained on all representative parts of the WSIs, in order to predict reliably at large. Biases may occur when human annotators discard parts which by their knowledge are irrelevant to a problem.
For the purpose of this experiment, we trained another model to discriminate between cancerous and non-cancerous tissue while this time we excluded necrotic tissue samples from the SKCM dataset. Necrosis is indeed a histopathological structure that is frequently observable in histopathological images but which is not related to the discrimination task. We then test the classifier on necrotic samples and compare it to a classifier which was trained on the entire SKCM dataset (including necrotic tissue samples).\\
Heatmaps of both models on an exemplary necrosis tile are depicted in the bottom row of Fig.~\ref{fig:bias_summary} (for more samples please refer to Supplemental Fig.~9). The heatmaps illustrate that large parts of the necrosis are mistaken for cancer when necrosis was not included in the training data (left heatmap) while the model trained on the comprehensive dataset correctly distinguishes the cancerous regions from healthy tissue independent of necrosis. We evaluated this in more depth on five tiles containing one to five, differently sized areas of necrotic tissue (Supplemental Fig.~9). Comparing the mean over these regions between the two classifiers, we observe that the biased training set leads to positive relevance on necrotic regions for half of the considered regions (cf. Supplemental Fig.~8). One can furthermore observe that the mean relevance per area of the unbiased classifier is rather constant, while there are high fluctuations between mean relevances of the biased classifier.

\section*{Discussion}
\label{sec:discussion}

\subsection*{Benefits of visual explanations}
By computing high-resolution heatmaps we were able to take the evaluation of the model from patch-level to the relevant level of cells. This is under the assumption that a morphological classifier would base its decision on all occurrences of the corresponding cell type in order to ensure reliable predictions on the very heterogeneous WSIs. In morphological decisions, context also plays a central role as not only a cell's morphology by itself but also adjacent tissue or the relative positions of surrounding cells can be relevant for discrimination. Therefore, patch sizes are usually chosen to be in a range where they include multiple cells and other morphological structures. However, we would like to emphasise that a more detailed level of evaluation permits to add detailed insights on the cell level. 
Here, cell-level resolution is achieved by averaging over single pixels within a circular region of a predefined radius around the given point annotations. Alternatively, one could also consider the use of explanation methods, which render heatmaps at the corresponding resolution, \eg by not fully backpropagating relevances to the input layer but instead stopping at arbitrary layers (\eg LRP). However, this has the disadvantage that for different cell-types one has to chose different resolutions while pixel-wise heatmaps are in this sense more flexible in resolution.  
Another theoretical possibility of tackling the challenge of cell localisation is to apply object detection algorithms from computer vision. In practice, however, this may not be feasible as it would require single cell annotations on a large data corpus. In the present analysis, we use these expensive single cell labels only for evaluating our approach, hence only requiring a small fraction of the otherwise required large amount of training labels.
With the ROC curves we showed that the computed heatmaps are indeed meaningful in the sense that they show a high overlap with labels provided by a board-certified pathologist. We examined the case where a classifier which performed well on patch-level, showed considerable shortcomings on the scale of cells (cf. SKCM in Fig.~\ref{fig:ROC}). Here, we observed that small but positive relevance values were also attributed to tumour-infiltrating lymphocytes (TILs), which were present in two (out of five) tiles. Disregarding these tiles, the AUC improves to 94\%. Hence, the evaluation on cell level led to an additional insight into the model. This helps to judge the model's suitability for a given task in digital pathology.
Additional to the validation aspect in the development and research phase, visual explanations also offer great potential for practical applications of deep learning-based analyses in digital pathology routine diagnostics. In particular, visual explanations could, for example, serve in applications with experts-in-the-loop or to speed up approval processes for medical products that involve deep neural networks (because external reviewers can easily validate trained models).\\
The known contrary behaviour of precision and recall when considering different, fixed sampling ratios in mini-batches is also reflected in the corresponding heatmaps (cf. Fig. \ref{fig:sampling}). This has important implications for medical application as one can adjust the strategy depending on the application's goal, emphasising recall or precision accordingly. For example, in applications that support pathologists in the very time-consuming process of detecting micro metastases in large numbers of lymph nodes, high sensitivity is the main criterion, \ie we are interested in all cells and structures that might potentially be cancerous, even those with a low likelihood.\\ 
The second main beneficial aspect of explanation heatmaps concerns the revelation of undesired strategies of the model, which are provoked by unnoticed biases in the data (cf. Sec.~\emph{Experiments (Uncovering biases)}). Not only do incorrect predictions come at a high cost in the medical field, but there is also a high susceptibility to biases due to the interdisciplinarity, and limitations of data collection in the field. In particular, we study the effects of three types of biases, which include biases in the entire dataset, biases correlated to a single class and sample biases, which can all be observed in histopathological images.
In digital pathology biases which affect the entire dataset mostly emerge from procedures of processing or labelling the dataset and are therefore specific to the particular dataset. 
In contrast, the class-correlated bias usually originates from sampling only a particular subset of the class. This effect is often enlarged in medical imaging due to small sample sizes. While in large datasets these biases marginalise, they are more prominent in small datasets and therefore might easily lead to undesirable decision strategies. Examples from digital pathology include artefacts such as tissue folds or tears that could correlate with one of the classes by chance. Another example for a potential issue is the co-occurrence of tumour and TILs. If most of the samples indeed display this interaction, the classifier may learn to associate TILs with positive class labels instead of detecting cancer cells. This effect will be even more reinforced if the negative class rarely contains lymphocytes. Therefore there is an increased risk of falsely basing the decision only on the presence of lymphocytes.
Another main source of biases in digital pathology arises from sampling small amounts of data from large, very heterogeneous unlabelled image corpora. This collection of samples should ideally be representative for WSIs to allow for the application of the models to the latter. In this work, we illustrate this by neglecting necrotic tissue which is of frequent occurrence in histopathology but not directly related to detecting tumour cells.\\
In general, these biases may not be discovered using standard quantitative performance measures on patch level. In the following, we discuss the superiority of applying explanation methods in the detection of the three types of biases described above.
Models trained on a dataset which is entirely affected by a bias will perform well on test data (containing this bias) in terms of standard performance measures. In order to detect this type of bias with performance measures, one needs an independent test dataset (inherently not containing this bias), requiring another reasonably large, annotated dataset which is very time-consuming. Besides this need of a high number of samples necessary to recognise the misleading pattern, it is also inevitable to carefully manually inspect all (falsely) classified patches. With explanation heatmaps, however, we can detect inappropriate strategies, for example of focusing on the centre, either on a single patch of the test data or at the average over absolute heatmap values (cf. Fig.~\ref{fig:bias_summary} and Supplemental Fig.~5). 
Biases that are correlated to labels of a certain class are similarly difficult to be discovered with standard performance measures. Analogous to our findings, this has also been demonstrated before in a standard computer vision benchmark dataset where copyright tags were correlated with a specific class label \cite{lapuschkin2016analyzing}. Identifying that a certain artefact or biological feature is highly correlated with one of the class labels is extremely difficult and demands visual inspection of a large number of samples. Not only can domain knowledge be necessary, \eg in the case of correlated biological features, but also features or artefacts might be so small that in certain resolutions they can even be invisible to the human eye. High-resolution explanation heatmaps help to identify this type of bias on a single affected heatmap.
Visual inspection of held-out data furthermore helps to identify missing sample classes on a single representative of this class despite a possibly small occurrence in the data. In contrary, standard performance metrics require a labelled test dataset and a significantly higher number of this missing sample class in order to be able to detect it. Furthermore, visual inspection allows to pinpoint the cause of misclassification precisely and trace it back to even small structures. This, however, only allows to find missing classes that are misclassified by the particular model and it is not generic for finding all missing classes.
It allows one to start with an initial dataset and augment it by those samples for which initial versions of the prototype perform poorly, resulting in an iteratively enriched dataset. Explainable machine learning allows to check large amounts of unlabelled image data for cases where the system performs poorly, with the aim to augment test and training data by asking domain experts to annotate only these failed cases and thus creating the next version of training and test set. 
In summary, visual explanations facilitate detecting biases on single, affected patches without the need of further large amounts of annotated data. \\

\subsection*{Advantage of pixel-wise heatmaps}
\begin{figure}[h!]
	\centering
	\includegraphics[width=0.7\textwidth]{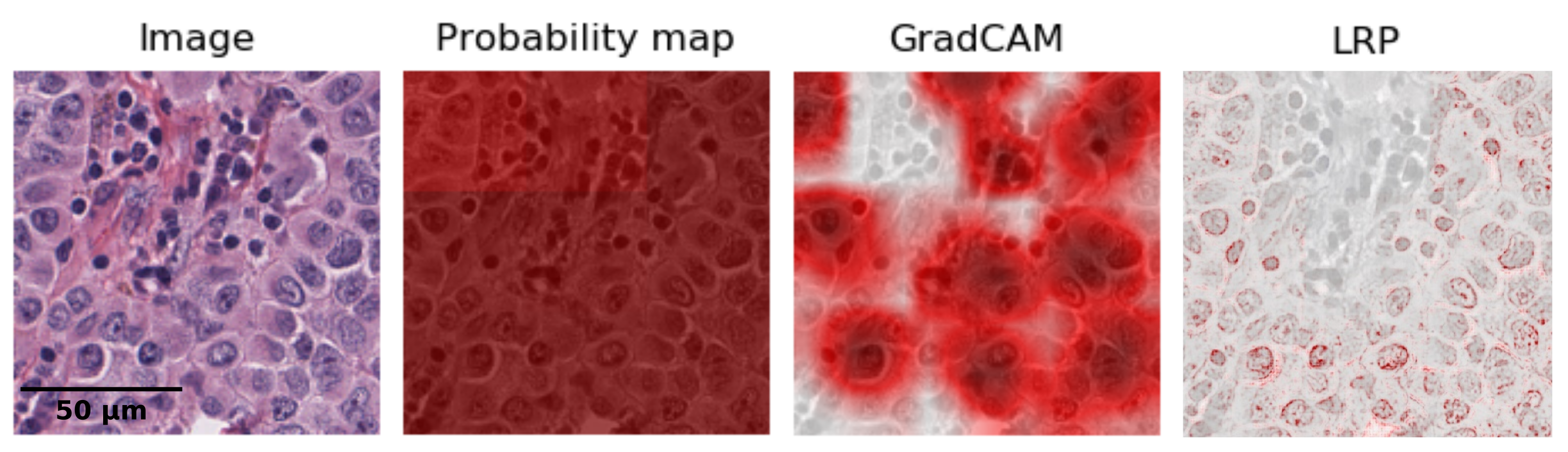}
	\caption{Different granularity levels of visual explanations---from patch level (middle left) to pixel-wise heatmaps (right). The H\&E image is split into 9 patches and explanation heatmaps for class \emph{cancer} are generated by different methods, namely probability map, Grad-CAM and LRP.}
	\label{fig:grad_cam}
\end{figure}

\noindent We like to remark that the presented benefits in this work are not limited to the choice of Layer-wise Relevance Propagation as explanation method. More generally, this should hold true for (fine-grained) heatmap approaches in general as the analyses are not based on any specific properties of the heatmap generation other than its high resolution. We argue that high-resolution heatmaps have the advantage (over rather low granularity) that even small structures can be spotted precisely, as otherwise, they could be smaller than the heatmaps' resolution. Therefore the high-resolution heatmaps provide very flexible data to be evaluated at various appropriate scales. 
For explaining decisions of neural networks in digital pathology, probability maps \cite{liu17} and Grad-CAM  \cite{korbar2017looking} have been proposed. Figure~\ref{fig:grad_cam} demonstrates these different levels of resolutions in heatmaps. The probability map is the output probability of the model per patch and is hence of the lowest possible native resolution. This leads to rather poor spatial resolution due to the patch size which considers the trade-off between localisation and context. The \textit{Grad-CAM} heatmaps, similar to Korbar et al.~\cite{korbar2017looking}, are more localised than the probability maps but only in the resolution of the last feature map. The corresponding ROC curves can be found in Supplemental Fig.~11. Similar resolution was obtained by heatmaps for support vector machines with Bag-of-Visual-Words Features~\cite{binder2018towards}, where the resolution is limited by the construction of the features, in particular, the grid size as well as the scale of local features.  
Although Korbar et al.~\cite{korbar2017looking} also render heatmaps in pixel space by using guided Grad-CAM, they, however, use the heatmaps to create rectangular regions around relevant pixels in order to measure the overlap of highlighted regions with expert-annotated regions of interest. Therefore they do not make direct use of the high-resolution properties of the heatmaps. 
Therefore, this work is the first, to the best of our knowledge, that explores the potential of pixel-level-resolution explanation heatmaps in digital pathology in regard to detecting biases.\\

\subsection*{Limitations of visual explanations}
\noindent As mentioned above, not only a cell's morphology but also adjacent tissue and spatial relations to other cells can be decisive to discriminate between tumour and healthy cells. This is where visual explanations are limited by the absence of an unambiguous depiction of cell composition that is intuitive for humans. Therefore, we only measure the relevance score on cells and not in surrounding tissue to determine the ROC curves.

\section*{Conclusion}
\label{sec:conclusion}

We have shown that the benefits of pixel-wise visual explanations in deep learning-based approaches are two-fold: Firstly, we can, both quantitatively and quantitatively, compare features that were learned in an end-to-end fashion against labels provided by domain experts, and secondly uncover latent biases in datasets.   
We showed that by computing pixel-wise heatmaps with Layer-wise Relevance Propagation, we obtained reasonable visualisations of the model's decision process in the three presented tumour entities, which show a high overlap with labels provided by domain experts. The overlap with labels was measured using ROC curves to evaluate the networks' performances on cell-level. This is under the assumption that an ideal model considers all tumour cells in its decision, when distinguishing between healthy and cancerous tissue. 
Additionally, we could show that these heatmaps reflect knowledge about training procedures, which are known from common performance measures such as precision and recall. This has important implications for medical application due to its flexible adjustment to emphasising recall or precision accordingly to the application's goal. This was demonstrated by varying the class sampling ratio in mini-batches. 
Secondly, explanation heatmaps enable us to detect various types of biases, which often hamper generalisation abilities of the models. 
In particular, we showed the detection of three types of implicit biases: Firstly, biases which affect the entire dataset, secondly, biases which are by chance correlated to a particular class label and thirdly sampling biases. With high-resolution heatmaps, uncovering these biases is possible on single samples and especially without the need of large, held-out labelled data corpora, contrary to standard performance measures. 
While uncovering these biases helps us to eliminate them from the data, heatmaps offer an important tool to iteratively update the dataset with missing sample classes.
In summary, high-resolution heatmaps present a versatile tool for both medical and machine learning professionals to gain insight into complex machine learning models in the deployment phase as well as in the research and development phase, respectively.

\bibliography{bibliography_patho}

\section*{Acknowledgements}
\noindent This work was supported by the German Ministry for Education and Research as Berlin Big Data Centre (01IS14013A) and Berlin Center for Machine Learning (01IS18037I \& 01IS18037A). Partial funding by DFG is acknowledged (EXC 2046/1, project-ID: 390685689). KRM is also supported by the Information \& Communications Technology Planning \& Evaluation (IITP) grant funded by the Korea government (No. 2017-0-00451). SL and WS are supported by BMBF TraMeExCo grant 01IS18056A. PS is supported by BMBF MALT III grant 01IS17058. AB was supported by Ministry of Education Tier2 Grant T2MOE1708 and the STEE-SUTD Cyber Security Laboratory. This publication only reflects the authors views. Funding agencies are not liable for any use that may be made of the information contained herein.

\newpage
\appendix
\setcounter{figure}{0}
\setcounter{table}{0}

\phantom{.}  
{\center \huge \bf Supplemental Document\par}
\vspace{1cm}

\appendixtitle{A}{\textbf{Confusion matrices}}

\begin{figure}[htb]
    \centering
    \includegraphics[width=.6\textwidth]{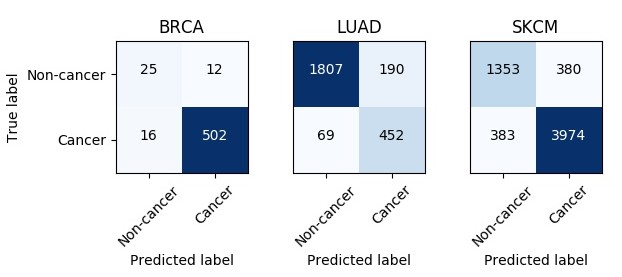}
    \caption{Confusion matrices of the classifiers discriminating between cancerous and healthy tissue in the three studied tumor entities, namely invasive breast cancer (BRCA), lung adenocarcinoma (LUAD) and cutaeneous malignant melanoma (SKCM). When considering these results, please keep the different label distributions of the datasets in mind.}
    \label{fig:conf_matrix}
\end{figure}

\appendixtitle{B}{\textbf{Detailed information on available annotations for evaluation on cell level}}

\begin{figure}[h!]
    \centering
    \includegraphics[width=.8\textwidth]{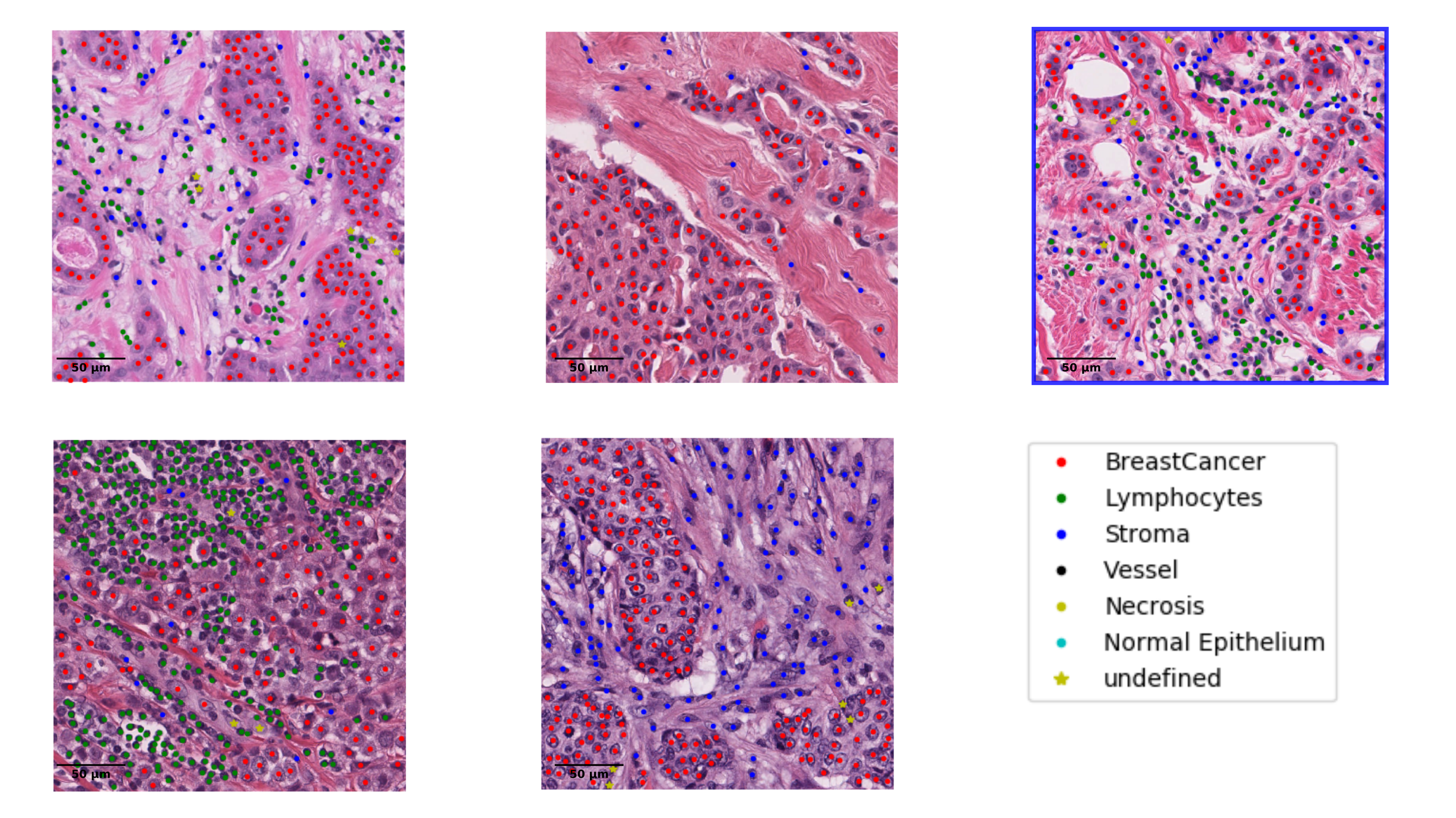}
    \caption{Chosen exemplary sample tiles from the BRCA project, superimposed with the extensive single cell annotations. The blue box marks the example shown in the paper.}
    \label{fig:eval_brca_samples}
\end{figure}

\begin{figure}[h!]
    \centering
    \includegraphics[width=.8\textwidth]{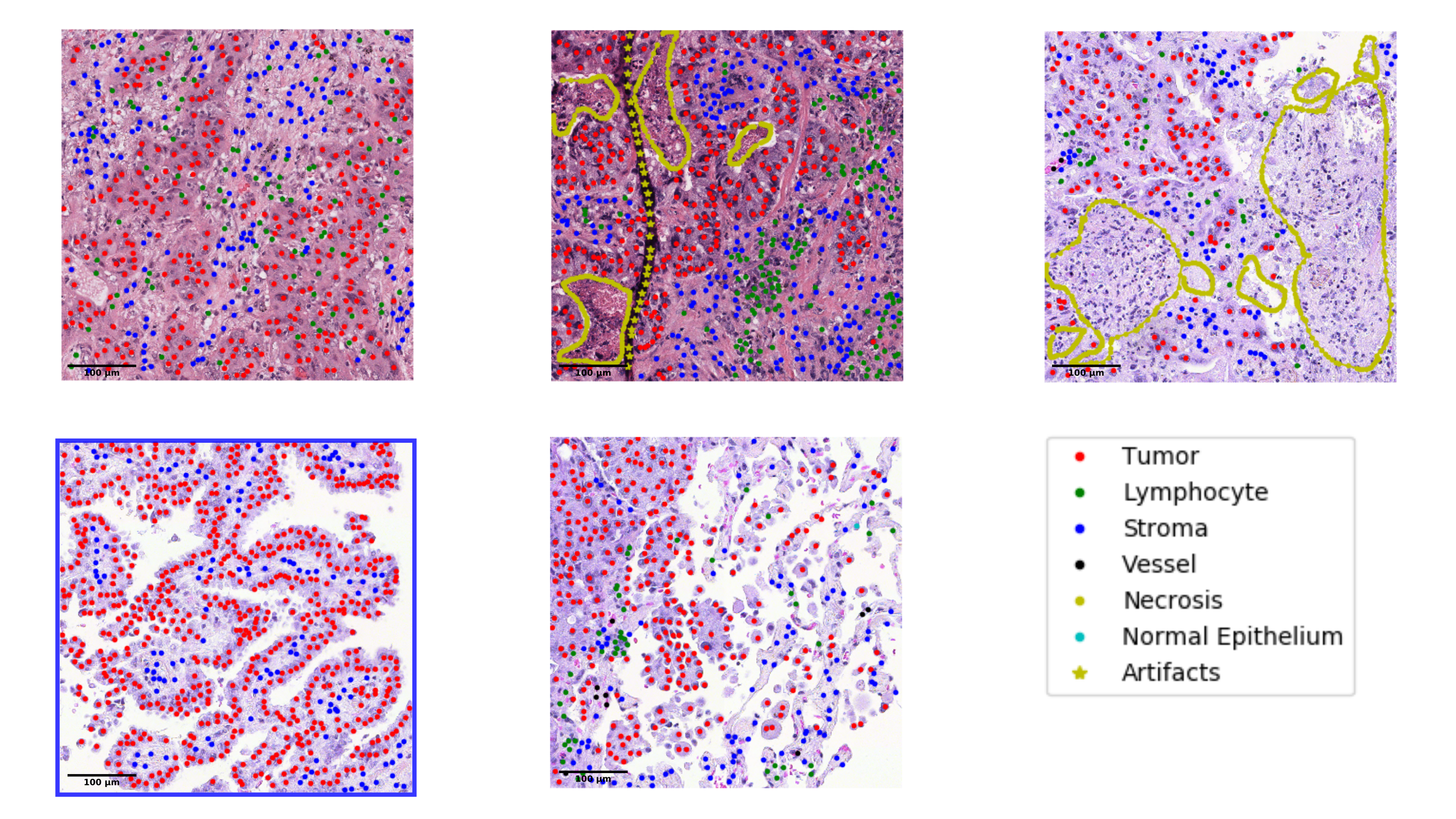}
    \caption{Chosen exemplary sample tiles from the LUAD project, superimposed with the extensive single cell annotations. The blue box marks the example shown in the paper.}
    \label{fig:eval_luad_samples}
\end{figure}

\begin{figure}[h!]
    \centering
    \includegraphics[width=.8\textwidth]{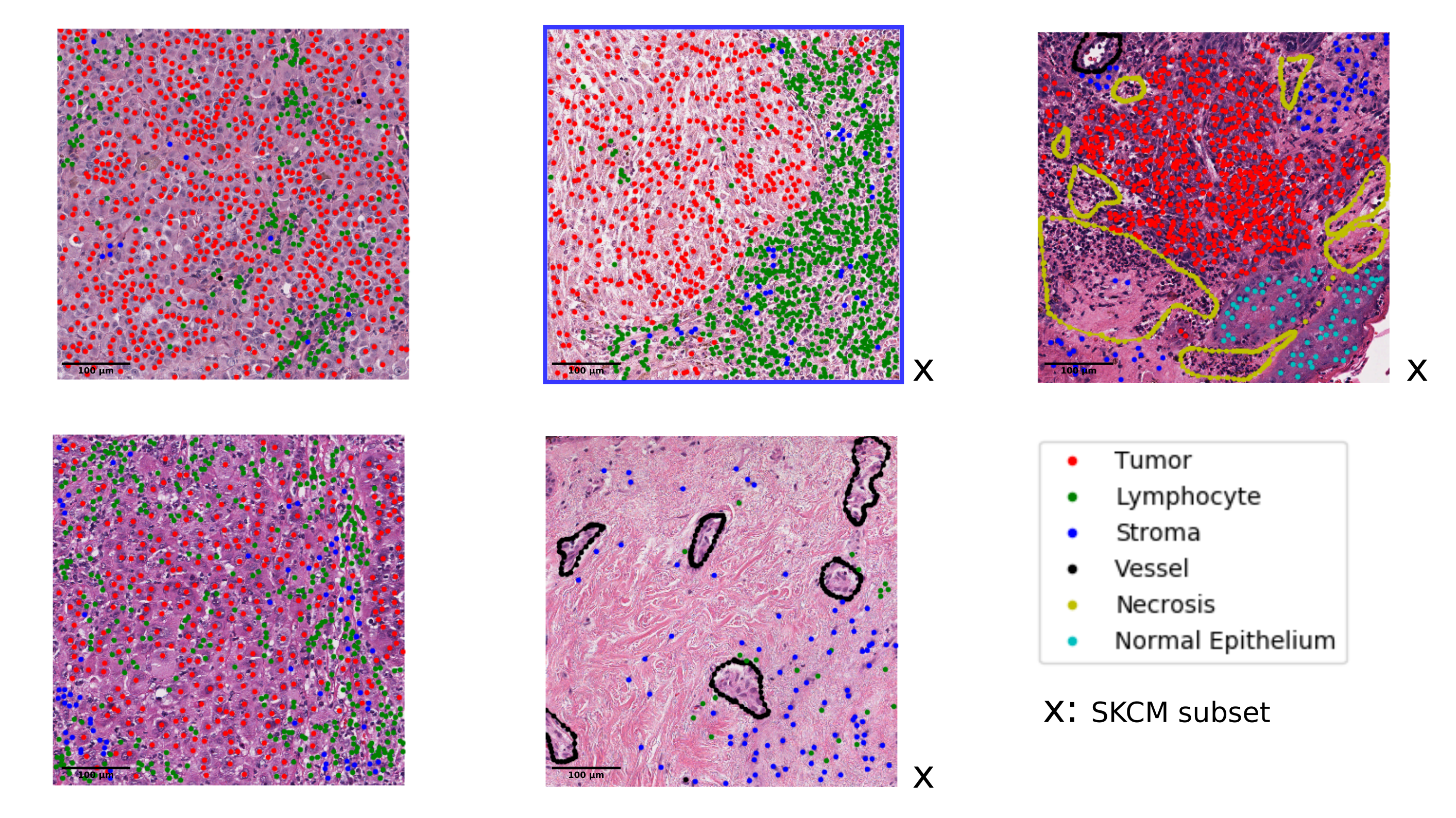}
    \caption{Chosen exemplary sample tiles from the SKCM project, superimposed with the extensive single cell annotations. The blue box marks the example shown in the paper. The black marks show the subset which excludes the two tiles that contain tumour-infiltrating lymphocytes.}
    \label{fig:eval_skcm_samples}
\end{figure}

\begin{table}[h!]
	\centering
	\caption{Summary of the available annotations for evaluation on cell level.}
	\begin{tabular}{lrr}
	\toprule
	 Tumor entity & \multicolumn{1}{c}{Total number} & \multicolumn{1}{c}{Number of} \\
	  & \multicolumn{1}{c}{of cells} & \multicolumn{1}{c}{cancer cells} \\
	 \midrule
	 \Tstrut 
	 BRCA 			& 1,803 & 820 \\
	 SKCM 		    & 3,961 & 2,247 \\
	 LUAD 		    & 2,722 & 1,650 \\
	 \bottomrule
	\end{tabular} 
	\label{tab:data_ROC}
\end{table}

\FloatBarrier
\pagebreak
\appendixtitle{C}{\textbf{Data sampling strategies}}

\begin{table}[htb]
    \caption{Performances of models with different sampling ratios.}
    \centering
    \begin{tabular}{cccc}
    \toprule
	 Sampling ratio & $F_1$-score & Recall (\emph{cancer} class) & Precision (\emph{cancer} class)\\
	 \midrule
	 \Tstrut 
	 0.5    & 0.92 & 0.88 & 0.97 \\
	 0.8	& 0.93 & 0.92 & 0.95 \\
	 \bottomrule
    \end{tabular}
    \label{tab:sampling_ratio}
\end{table}

\vspace{1cm}
\appendixtitle{D}{\textbf{Uncovering biases}}

\begin{center}
	I. Dataset bias
\end{center}

\begin{figure}[h!]
    \centering
    \includegraphics[width=.55\textwidth]{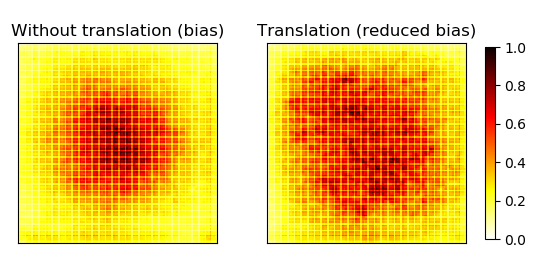}
    \caption{Mean absolute relevances (of 896) patches of the "biased" (left) and the "unbiased" classifier (right). As expected relevance is predominately distributed in the center of the heatmap for the biased classifier.}
    \label{fig:centerbias_av}
\end{figure}

\begin{figure}[h!]
    \centering
    \includegraphics[width=.55\textwidth]{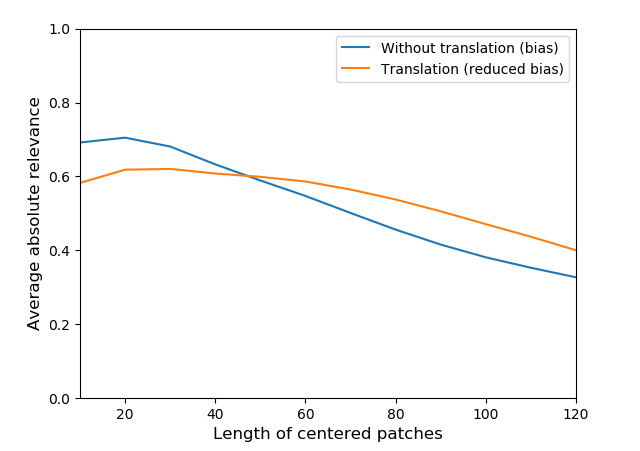}
    \caption{Evaluation of the effect of training on center biased data. We compare the mean amount of absolute relevance as a function of relative, centered areas on an independent (not affected) test dataset.}
    \label{fig:centerbias_relevance}
\end{figure}

\begin{figure}[h!]
    \centering
    \includegraphics[width=.55\textwidth]{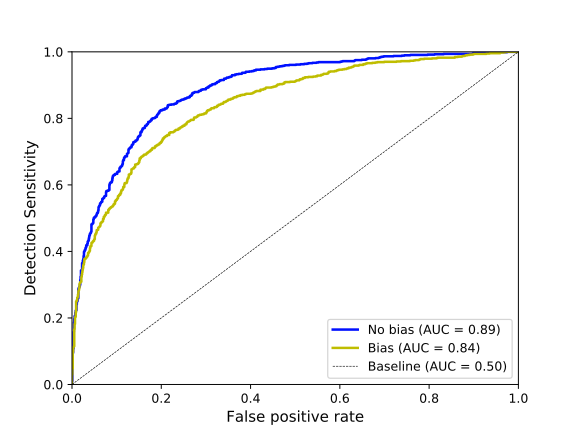}
    \caption{Receiver operating characteristic (ROC) curve on seven held-out, extensively labelled 1000x1000px tiles of the corresponding dataset. The ROC curve improves when counteracting the bias by random translations.}
    \label{fig:centerbias_ROC}
\end{figure}

\FloatBarrier

\vspace{2cm}
\begin{center}
	II. Sampling bias
\end{center}

\begin{figure}[h!]
    \centering
    \includegraphics[width=.6\textwidth]{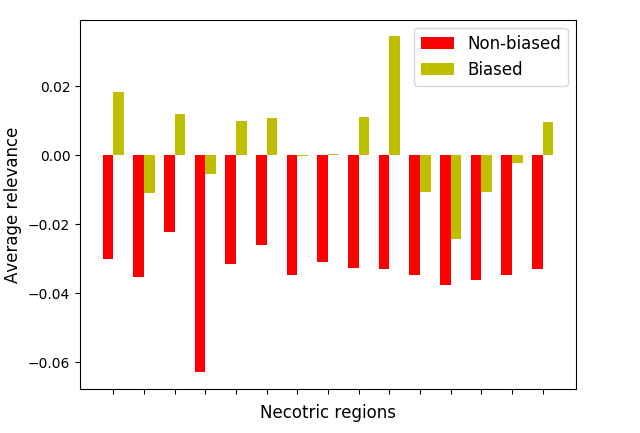}
    \caption{Evaluation of all annotated necrosis regions on exemplary necroses tiles (cf. Fig.~\ref{fig:necroses}). Comparison between the average relevance per region depending on the classifier.}
    \label{fig:eval_necrosis}
\end{figure}

\begin{figure}[h!]
    \centering
    \includegraphics[width=.8\textwidth]{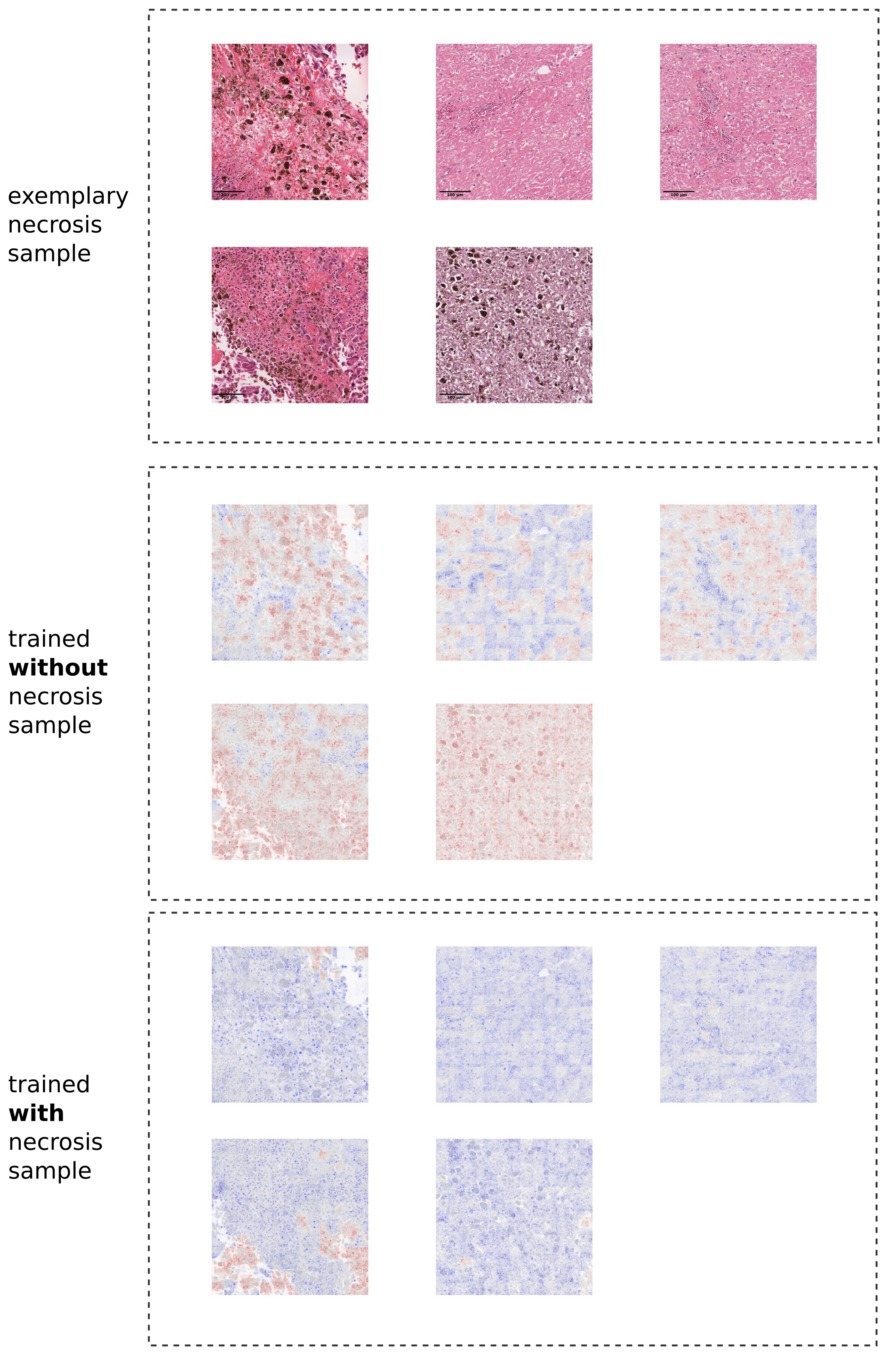}
    \caption{Exemplary H\&E tissue samples of necroses as well as heatmaps for both classifiers, \ie trained on a comprehensive dataset and trained on a dataset lacking necroses samples respectively.}
    \label{fig:necroses}
\end{figure}
\FloatBarrier

\vspace{2cm}
\begin{center}
	III. Class correlated bias
\end{center}

\begin{figure}[h!]
    \centering
    \includegraphics[width=.6\textwidth]{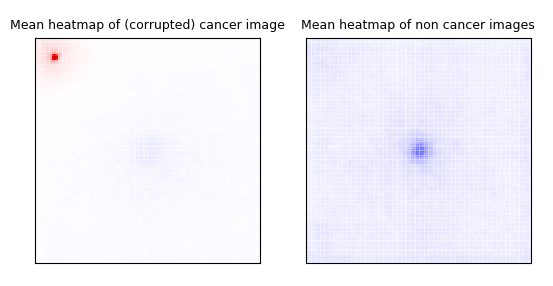}
    \caption{Average heatmaps for class \textit{cancer} of images classified as cancer (left) or classified as non-cancer (right).}
    \label{fig:correlated_bias_av}
\end{figure}

\FloatBarrier

\appendixtitle{E}{\textbf{ROC Curve for GradCam}}
\begin{figure}[htb]
    \centering
    \includegraphics[width=.6\textwidth]{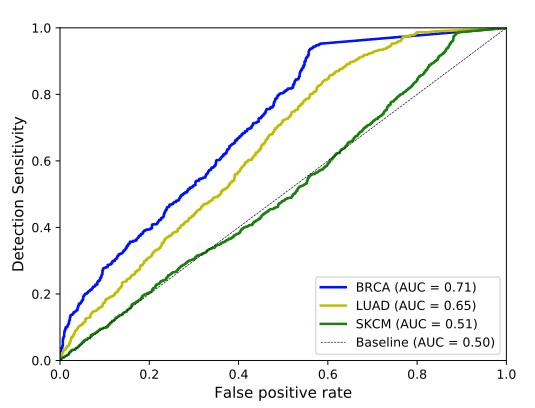}
    \caption{Receiver operating characteristic (ROC) curves for GradCam heatmaps on all three studied tumor entities.}
    \label{fig:gradcam}
\end{figure}

\FloatBarrier

\end{document}